\PassOptionsToPackage{english}{babel}   
\documentclass{volcanica-template}
\let\cite\supercite
\nonstopmode
\nolinenumbers
\usepackage{xspace}
\usepackage{amsmath}
\usepackage{mathtools,bm}
\usepackage{booktabs,longtable,array}
\usepackage{tikz}
\usetikzlibrary{arrows.meta,positioning}
\newcommand{\vect}[1]{\bm{#1}}

\DeclareRobustCommand{\craft}{CRAFT-$\varepsilon$\xspace}   

\newcommand{\ArticleType}{Research article}

\newcommand{\Title}{
CRAFT-$\varepsilon$: A Conformer-Resolved Atom- and Fragment-based Theory for the Static Relative Permittivity of Pure Organic Liquids
}

\newcommand{\shortTitle}{CRAFT-$\varepsilon$: Permittivity of Pure Liquids from Conformer Ensembles}
\newcommand{\Author}{M\"uller,}

\author[{{\affiliation{1}}}]
{\orcidaffil{0000-0003-1684-6994}~
Simon M\"uller}
\affil[{{\affiliation{1}}}]{
Institute of Thermal Separation Processes,
Hamburg University of Technology,
Hamburg, Germany
}

\begin{document}

\FrontMatter{
The static relative permittivity is an important property in thermodynamic modeling, electrochemistry, and solvent design, but predicting it directly from molecular structure remains challenging because both electronic polarization and molecular dipole orientation distribution in the liquid phase contribute to the measured response. Here, \craft (Conformer-Resolved Atom- and Fragment-based Theory for $\varepsilon$) is introduced, a hybrid semi-empirical, conformer-resolved molecular descriptor model for predicting the static relative permittivity of neutral pure liquids. Conformer free energies, molecular dipoles, and atom-resolved polarizabilities are obtained from a GFN2-xTB/ALPB workflow. The molecular dipole is decomposed into chemically defined polar-fragment vectors, allowing fitted corrections to retain both the magnitude and relative orientation of local dipolar contributions. Electronic and orientational polarization are combined with a group-contribution molar volume in the Onsager--Kirkwood--Fr\"ohlich relation. A staged parameterization with 41 parameters is applied to nonpolar, polar, and hydrogen-bonding compounds. For 1393 liquids with permittivities between 1.5 and 183, \craft reaches a five-fold cross-validated average absolute deviation of 2.40 ($R^2=0.93$). Conformational averaging is essential for liquids with several polar groups. Compared with an established fragment-based model with 139 parameters, it reaches a similar cross-validated RMSE (3.65 versus 3.41) with about a quarter of the parameters and applies to a broader set of liquids due to the more general character of the approach. The resulting framework provides a computationally inexpensive route from molecular structure to liquid dielectric response and forms a basis for later application to solvent mixtures and electrolyte systems.
}
[]%
{
\vspace{1.5\baselineskip}
{Relative permittivity}
{Molecular descriptors}
{Polarizability}
{Dipole moment}
{Conformer ensemble}
}


\section{Introduction}\label{introduction}

The static relative permittivity, $\varepsilon_r$, is one of the fundamental macroscopic properties connecting molecular-scale electrostatics with the behavior of condensed phases. It determines the screening of electrostatic interactions and consequently enters descriptions of solvation, electrolyte thermodynamics, reaction media, phase equilibria, and electrochemical systems. Reliable values are required both as experimental material properties and as input quantities for predictive thermodynamic models.

For electrolyte systems in particular, the relative permittivity directly
affects long-range electrostatic contributions such as Debye--H\"uckel-type terms \cite{debye_zur_1923} as well as Born-type descriptions of ionic solvation \cite{born_volumen_1920}. Predictions of ionic activity coefficients and phase equilibria can consequently be highly sensitive to the dielectric model employed \cite{muller_calculation_2020,rueben_permittivity_2024}. In practical applications, experimental dielectric constants or empirical correlations are frequently supplied externally to thermodynamic models. This limits predictivity when new solvents or solvent mixtures are considered.

Predicting $\varepsilon_r$ from molecular structure is challenging because the static dielectric response contains physically distinct contributions. The rapid electronic response is related primarily to molecular polarizability, while the usually larger low-frequency contribution of polar liquids originates from molecular reorientation in the liquid. The latter depends not only on the isolated-molecule dipole moment but also on molecular conformation, local liquid structure, intermolecular association, and dipole--dipole correlations. Classical dielectric theories relate both contributions to molecular properties: the Clausius--Mossotti relation connects the electronic polarizability with the high-frequency response, whereas the theories of Onsager, Kirkwood and Fr\"ohlich describe the orientational polarization of permanent dipoles, with Onsager accounting for the local field through a cavity reaction field and Kirkwood and Fr\"ohlich additionally introducing dipole--dipole correlations through the Kirkwood correlation factor \cite{bottcher_theory_1973,onsager_electric_1936,kirkwood_dielectric_1939,frohlich_theory_1958}. Their direct use as predictive models is nevertheless difficult, because effective liquid-phase dipoles, local-field effects and orientational correlations are generally not known from molecular structure alone, and a single optimized gas-phase geometry discards conformational information that can be relevant in the liquid.

An alternative is to retain these physically motivated dielectric relations while representing the missing condensed-phase effects through transferable molecular descriptors and a limited number of fitted parameters. Such a model is hybrid and semi-empirical. While the functional form follows dielectric theory and the molecular inputs are computed from the structure, additionally fitted effective scale factors account for the liquid-phase effects that the theory and the isolated-molecule calculations do not capture. Such an approach may be particularly useful for high-throughput applications, where large numbers of candidate molecules must be evaluated at a computational cost that prohibits explicit condensed-phase sampling.


\section{State of the Art}
\label{chap:state-of-the-art}

\subsection{Permittivity in electrolyte models}

Electrolyte solutions are commonly described within two frameworks \cite{reimer_primitive_2025}. Non-primitive models represent the solvent molecules explicitly and account for the interactions between ions and solvent molecules. Primitive models treat the solvent as a dielectric continuum, characterized only by its relative permittivity, and represent only the ions explicitly, for example through the Debye--H\"uckel equation \cite{debye_zur_1923} and the mean spherical approximation (MSA) \cite{waisman_exact_1970,blum_mean_1975} for ion--ion interactions, and the Born theory of solvation \cite{born_volumen_1920} for ion--solvent interactions. A comparison with molecular simulations of non-primitive model electrolytes showed that the primitive-model expressions provide a qualitatively reasonable description, but systematically underestimate the Helmholtz energy contributions of ion--solvent and ion--ion interactions unless the Born radii are adjusted empirically \cite{reimer_primitive_2025}.

Non-primitive approaches instead require models for the dipolar solvent and its interactions with the ions. Perturbation theories have been developed for Stockmayer fluids on the basis of dipolar hard spheres \cite{theiss_equation_2019} and extended to dipolar fluids and electrolyte solutions by splitting the electrostatic pair potentials into short- and long-ranged contributions. These theories are considerably more accurate than the MSA but show limitations for partial molar quantities such as the mean activity coefficient \cite{theiss_perturbation_2020}. Helmholtz energy models for dipole interactions have been reviewed and assessed comprehensively \cite{staubach_helmholtz_2024}. Perturbation theories for the relative permittivity of dipolar model fluids have been assessed against molecular simulation data \cite{kohns_critical_2021}, and a molecular theory based on a modified mean-field approximation has been compared with new simulation data for dipolar hard-sphere and Stockmayer fluids \cite{kournopoulos_molecular_2022}.

Most engineering models for electrolyte solutions, for example the Pitzer model \cite{pitzer_thermodynamics_1973}, electrolyte NRTL \cite{chen_local_1986}, extended UNIQUAC \cite{thomsen_modeling_1999}, ePC-SAFT \cite{held_modeling_2008} and SAFT-VRE \cite{galindo_saft-vre_1999}, however, retain a primitive-model description of the long-range electrostatics and therefore require the permittivity of the solvent or solvent mixture as an input. In the electrolyte model COSMO-RS-ES, which is based on the same framework as openCOSMO-RS \cite{gerlach_open_2022,grigorash_comprehensive_2025,grigorash_comprehensive_2026,joos_importance_2026,ahmed_fast_2026,markgraf_extension_2027}, the permittivity of the solvent or solvent mixture enters the long-range electrostatic contribution and thereby the calculated activity coefficients, salt solubilities and phase equilibria \cite{muller_calculation_2020}. Systems with low or composition-dependent permittivity, as found in organic and mixed solvents, require dedicated corrections of this contribution. \craft pursues a simple hybrid route based on semi-empirical quantum-chemical calculation to predict this important property.

\subsection{Molecular simulation and dipolar model fluids}

Molecular dynamics (MD) simulation provides the most direct molecular route to static relative permittivity, evaluating either equilibrium fluctuations of the total simulation-cell dipole moment or the response to an applied electric field. Methodological comparisons have established the importance of consistent electrostatic boundary conditions, adequate system sizes, and sufficiently long trajectories, while also showing that statistical uncertainty can remain substantial for slowly relaxing polar liquids like water \cite{riniker_calculation_2011,gereben_accurate_2011,kolafa_static_2014,sanchez_uncertainties_2019}. MD has consequently served both as a benchmark for dielectric theories and as a means to assess force fields against experimental data across broad sets of organic liquids, including fixed-charge, united-atom, and polarizable models \cite{caleman_force_2012,beauchamp_toward_2015,nunez-rojas_force_2018,liu_benchmarks_2018,barrera_new_2026,binninger_amoeba_2023}. These studies have shown that agreement for density or vaporization enthalpy does not ensure an accurate dielectric response: the effective liquid-phase dipole and  molecular polarizability must be represented consistently \cite{jorge_dielectric_2019,jorge_what_2023,jorge_theoretically_2024}. Simulations have further resolved the roles of hydrogen-bonded aggregation and temperature in carboxylic acids, liquid-phase dipole enhancement in alcohols, and ionic effects in salt-containing solvents \cite{riniker_temperature_2012,jorge_dipole_2022,shock_molecular_2023}. Thus, MD supplies a physically detailed reference for developing and testing theoretical closures \cite{kohns_molecular_2020,kournopoulos_molecular_2022,marx_co-oriented_2025}, but its computational cost and its dependence on the underlying interaction model limit its use as a routine high-throughput predictor. Even the one of the broadest force-field benchmarks cover on the order of 150 liquids \cite{caleman_force_2012}, far fewer than the chemical space relevant for solvent screening.

Complementary insight is obtained from MD studies of dipolar model fluids, where molecular shape, dipole strength, dipole location, density, and temperature can be varied independently. Simulations of Stockmayer and shifted-dipole fluids have provided controlled reference data for assessing dielectric theories and have demonstrated that orientational correlations, particularly in the first coordination shell, are in many cases not generally captured by simple perturbation approaches \cite{langenbach_relative_2020,kohns_critical_2021,kohns_molecular_2020}. This gap motivated the co-oriented fluid functional equation for electrostatic interactions (COFFEE), which represents the orientation distribution explicitly and has been extended to mixtures \cite{marx_co-oriented_2023}. The subsequent correction of the underlying simulation data also underlines the need for careful validation of dielectric calculations, especially for strongly dipolar fluids \cite{langenbach_correction_2020}.
Nevertheless, these theories remain impractical or even inadequate for complex multipolar molecules, particularly those containing several polar groups whose mutual interactions and relative orientations generate a heterogeneous local electrostatic environment.

Several recent studies have therefore attempted to improve how molecular models represent the polarization that governs the permittivity of real liquids. Jorge \textit{et al.} showed that the dipole moment of a fixed-charge model should lie approximately halfway between those of the gas and the liquid phase, and that the liquid-phase reference dipole must include both mean-field and local contributions to polarization. Dielectric continuum models capture only the former and therefore significantly underestimate the polarization of molecules in polar liquids \cite{jorge_what_2023}. The underlying halfway-charge and electronic-continuum theories have been reconciled into a force-field development workflow that combines an optimal model dipole with post facto polarization corrections \cite{jorge_theoretically_2024}, complementing the earlier correction of simulated dielectric constants for electronic polarization through the refractive index \cite{jorge_dielectric_2019}. The 2023 and 2024 studies identify the effective liquid-phase dipole as the key quantity and show that continuum descriptions of polarization alone are insufficient, which is the deficiency that the fitted fragment scales of \craft address empirically.

Analytical routes to the permittivity of real polar liquids have also been pursued. Rueben \textit{et al.} \cite{rueben_predicting_2023} used the third-order expansion of the perturbation theory for dipolar fluids, supplemented by a substance-specific correction for short-range correlations, to develop a group-contribution model that applies to a wide range of molecules. Morozov\cite{morozov_predicting_2026} later re-derived the same expansion without this correction for short-range correlations. The result is a universal expression, $\varepsilon = 1 + 3y_\mathrm{eff}\left(1 + y_\mathrm{eff} + y_\mathrm{eff}^2/16\right)$, as a function of an effective dipole density $y_\mathrm{eff}$, which combines the contributions of the permanent dipoles and the polarizability. It follows from the third-order virial series of dipolar hard spheres in which all terms depending on the short-range coupling parameter are eliminated \cite{morozov_predicting_2026}. It was applied to about one hundred polar liquids with permittivities up to 30 and extended to mixtures. As the dipole moments of molecules in the liquid are not known from the structure alone, experimental values, mainly from benzene solutions, were used as input. \craft instead computes the dipole moments and polarizabilities from the molecular structure and absorbs the neglected correlations in fitted effective scales, which allows it to be applied to liquids for which no experimental dipole is available.

\subsection{Data-driven structure--property models}

Quantitative structure--property relationships (QSPR) and machine-learning (ML) models provide a low-cost alternative to explicit simulation by learning the relation between permittivity and molecular descriptors such as composition, functional-group counts, connectivity, dipole moment, and polarizability. Early QSPR studies used multivariate regression on topological, geometrical, and quantum-chemical descriptors, and subsequent work applied descriptor selection and SMILES-based representations to broader organic and $\pi$-conjugated compound sets \cite{cocchi_development_1999,sild_general_2002,liu_quantitative_2010,lee_elucidation_2012,achary_qspr_2014}. Neural-network QSPR models demonstrated the same strategy for chemically diverse compounds and showed that chemically homogeneous subsets improve accuracy relative to a single global model \cite{schweitzer_development_1999,schweitzer_improved_2000}. More physically restricted approaches have combined simple molecular properties with dielectric relations for nonpolar hydrocarbons, while quantum-chemical calculations have been used directly for specialized classes such as energetic materials or solids only \cite{panuganti_simple_2016,robitaille_dielectric_2021}.

More recent studies combine ML with interpretable descriptors: Abraham solvation parameters and SHAP analysis have been used for pure organic liquids and water mixtures, and feature-selection workflows have produced explicit prediction equations, including for strongly polar materials \cite{deng_dielectric_2022,luo_physical_2025}. Representation-learning and hybrid methods aim to reduce the reliance on manually selected descriptors: three-dimensional molecular representations pretrained on large structural data sets can be fine-tuned for bulk-property prediction, whereas physics-informed neural networks combined with symbolic regression seek compact, physically meaningful correlations \cite{ou_high-accuracy_2025,luo_physical_2025}. All data-driven predictions nevertheless remain constrained by the quality and chemical coverage of their training data, and their extrapolation is limited when liquid-phase association, dipole--dipole correlations, or chemical functionalities are underrepresented. Isolated-molecule representations, moreover, do not capture liquid-phase conformational populations, hydrogen-bond networks, or long-range orientational correlations. This motivates descriptors that retain a direct connection to the physical contributions governing the dielectric response, combined with a physically based dielectric closure.

\subsection{Group-contribution methods}

Group-contribution methods estimate the permittivity from the identities and counts of predefined molecular fragments. Sheldon \textit{et al.} correlated dielectric constants and related properties with UNIFAC group counts, offering a simple procedure that requires no three-dimensional molecular structure \cite{sheldon_pure_2005}. Bouteloup and Mathieu combined fragment contributions with the Kirkwood--Fr\"ohlich relation, representing the orientational dipolar term through additive polar-fragment parameters and obtaining a model applicable over a wide range of polarity, at the cost of many fitted parameters \cite{bouteloup_predicting_2019}. More recently, Rueben \textit{et al.} coupled perturbation theory with a group-contribution description to predict the temperature-dependent permittivity of pure liquids, with an extension to mixtures that was demonstrated for a few binary systems \cite{rueben_predicting_2023}. Such methods are computationally efficient and chemically interpretable, but simple group counts cannot distinguish conformations, the relative orientation of multiple polar fragments, or changes in local association caused by their proximity.

\begin{longtable}{@{}p{0.10\textwidth}p{0.25\textwidth}p{0.09\textwidth}p{0.26\textwidth}p{0.15\textwidth}@{}}
\caption{Overview of ML, QSPR, and group-contribution approaches for predicting relative permittivity.}
\label{tab:literature_ml_qspr_gc}\\
\toprule
Approach & Reference & Dataset size & Molecular scope & Permittivity range around \SI{25}{\degreeCelsius}\\
\midrule
\endfirsthead

\toprule
Approach & Reference & Dataset size & Molecular scope & Permittivity range around \SI{25}{\degreeCelsius}\\
\midrule
\endhead

\bottomrule
\endfoot

QSPR &
Cocchi et al., 1999 \cite{cocchi_development_1999} &
$\sim 40$ & varied organic molecules & $\sim 2$--$40$ \\

QSPR &
Sild \& Karelson, 2002 \cite{sild_general_2002} &
$\sim 200$ & varied organic molecules & $\sim 2$--$47$ \\

QSPR &
Liu et al., 2010 \cite{liu_quantitative_2010} &
$\sim 690$ & mainly single polar group & $\sim 2$--$50$ \\

QSPR &
Lee et al., 2012 \cite{lee_elucidation_2012} &
$\sim 150$ & $\pi$-conjugated compounds with mainly single polar group & $\sim 2$--$45$ \\

QSPR &
Achary, 2014 \cite{achary_qspr_2014} &
$\sim 150$ & $\pi$-conjugated compounds with mainly single polar group & $\sim 2$--$35$ \\

ML &
Schweitzer \& Morris, 1999 \cite{schweitzer_development_1999} &
$\sim 500$ & varied organic molecules & $\sim 1$--$40$ \\

ML &
Schweitzer \& Morris, 2000 \cite{schweitzer_improved_2000} &
$\sim 450$ & varied organic molecules & $\sim 1$--$40$ \\

ML &
Deng \& Jia, 2022 \cite{deng_dielectric_2022} &
$\sim 30$ & mainly single polar group & $\sim 2$--$80$ \\

ML &
Mao et al., 2022 \cite{mao_development_2022} &
$\sim 1200$ & varied organic molecules & $\sim 2$--$180$ \\

ML &
Ou et al., 2025 \cite{ou_high-accuracy_2025} &
$\sim 950$ & mainly single polar group & mainly $\sim 2$--$60$ \\

ML &
Luo et al., 2025 \cite{luo_physical_2025} &
$\sim 30$ & mainly single polar group & $\sim 2$--$80$ \\

Group \mbox{contribution} &
Sheldon et al., 2005 \cite{sheldon_pure_2005} &
$\sim 340$ & mainly single polar group & mainly $\sim 2$--$50$ \\

Group \mbox{contribution} &
Rueben et al., 2023 \cite{rueben_predicting_2023} &
$\sim 260$ & single polar group & $\sim 2$--$180$\\

Group \mbox{contribution} &
Bouteloup \& Mathieu, 2019 \cite{bouteloup_predicting_2019} &
$\sim 1200$ & varied organic molecules & $\sim 2$--$180$ \\

\end{longtable}

\subsection{Research gap and objectives}

As summarized in Table~\ref{tab:literature_ml_qspr_gc}, many earlier QSPR and ML studies were developed using relatively small, curated data sets of simple organic liquids, frequently dominated by molecules with one principal polar functionality and by permittivities below $\varepsilon_r \approx 50$. The broadest exceptions are the fragment-based Kirkwood--Fr\"ohlich model of Bouteloup and Mathieu and the data-driven model of Mao \textit{et al.}, which are based on the same dataset. This dataset covers diverse pure liquids with values up to $\varepsilon_r=180$ \cite{mao_development_2022,bouteloup_predicting_2019}.

The purpose of this work is to evaluate how far a physically motivated model, based on inexpensive quantum-chemical calculations of conformer ensembles, can predict the static relative permittivity of neutral pure liquids without explicitly representing the microscopic structure of the fluid. Since directional intermolecular interactions are not treated explicitly, hydrogen-bonding liquids, particularly molecules containing multiple donor or acceptor sites, are expected to be more challenging than nonpolar compounds or molecules containing a single polar group. The model, referred to as \craft (Conformer-Resolved Atom- and Fragment-based Theory for $\varepsilon$), is conceptually related to that of Bouteloup and Mathieu \cite{bouteloup_predicting_2019}, but requires substantially fewer empirical parameters. Rather than introducing separate parameters for different configurations of the same polar functional groups, the present approach applies transferable scaling factors to molecular polarizabilities and dipole moments obtained directly from quantum-chemical calculations.

Instead of representing the molecular dipole by a single scalar quantity, the dipole of every conformer is decomposed into vectors associated with chemically defined polar fragments. These fragment dipoles can be scaled independently while retaining information about both their magnitudes and their relative orientations within the molecule. Electronic polarization is treated separately through atom-resolved polarizabilities. Both enter the Onsager--Kirkwood--Fr\"ohlich relation, and the model is parameterized in stages for nonpolar, polar, and hydrogen-bonding liquids in order to distinguish contributions associated with electronic polarization, permanent molecular dipoles, and directional intermolecular interactions.

The objectives of this work are therefore:
\begin{enumerate}
\item to establish a practical computational workflow connecting molecular structure, conformer ensembles, polarizability, and molecular dipoles to static relative permittivity;
\item to identify the chemical classes for which explicit condensed-phase structure remains necessary.
\end{enumerate}


\section{Methods}\label{methodology}

\subsection{Dataset}
\label{sec:dataset}

The dataset comprises static relative permittivities of 1393 pure liquids, compiled from two sources:
experimental values from the data collection of \citet{bouteloup_predicting_2019} (BM) and values obtained from the substance-specific perturbation-theory correlation of \citet{rueben_predicting_2023} (PM). For the PM data, the permittivity is evaluated with the liquid density from the corresponding density correlation. The temperature is 298.15\,K or, where this lies outside the range of validity, the nearest temperature within the range from the lower limit of the correlation to the smaller of $0.9\,T_\mathrm{c}$ and the upper limit of the density correlation. Of the 1393 liquids, 699 are covered only by BM, 210 only by PM and 484 by both sources. If all values of a compound lie within 10\,K of each other, they are averaged; otherwise, the value measured closest to 298.15\,K is used.

The records of both sources were merged on the basis of their chemical structures: names and CAS numbers were resolved to structures and cross-checked across multiple chemical databases with MoleculeResolver \cite{muller_how_2025}, and all structures are represented as canonical SMILES. Each record was additionally checked for consistency between its name and its structure: amide tautomers replace imidic acid forms, and the $E/Z$ configuration and relative ring stereochemistry implied by the name were assigned explicitly. Compounds with fewer than two carbon atoms (for example methanol or nitromethane) were removed, because in these molecules the polar group is not separated from the rest of the molecule by an alkyl chain. Their polar groups behave differently from those in larger molecules, so the transferable dipole scales of \craft do not describe them as well, and the group-contribution molar volume is less reliable for them for the same reason. Moreover, experimental permittivities are available for most of these small molecules, so a predictive model is less needed for them. In addition, a list of light compounds that are gaseous at room temperature and pressure (for example ethane, propane, butane, dimethyl ether and acetaldehyde) was removed. Compounds with a boiling point below room temperature that are not on this list, such as octafluoropropane, are retained; their PM values refer to the saturated liquid. Six further entries were excluded individually (details, including the reported values, in Table~S4 of the Supporting Information):
\begin{itemize}
  \item tetramethyl orthosilicate, whose value is inconsistent with the homologous silicate series;
  \item three hydrocarbons whose permittivities would require physically implausible dipole moments of
        1.0--1.7\,D;
  \item acetamide, which is a solid at the evaluation temperature;
  \item 2-pyrrolidone, the only lactam with an N--H group in the dataset, which is not sufficient to parameterize this polar group in the hydrogen-bonding stage as with only one molecule it would fit perfectly.
\end{itemize}

The liquids contain the elements C, H, O, N, S, P, Si, B, F, Cl, Br and I. Oxygen occurs in 869 molecules, nitrogen in 255 and halogens in 243. Molecular size ranges from 3 to 63 heavy atoms (median 9). The temperatures of the selected values range from 153 to 490\,K, and 1228 values (88\,\%) lie within 10\,K of 298.15\,K.

\begin{figure}[htbp]
  \centering
  \includegraphics[width=0.8\textwidth]{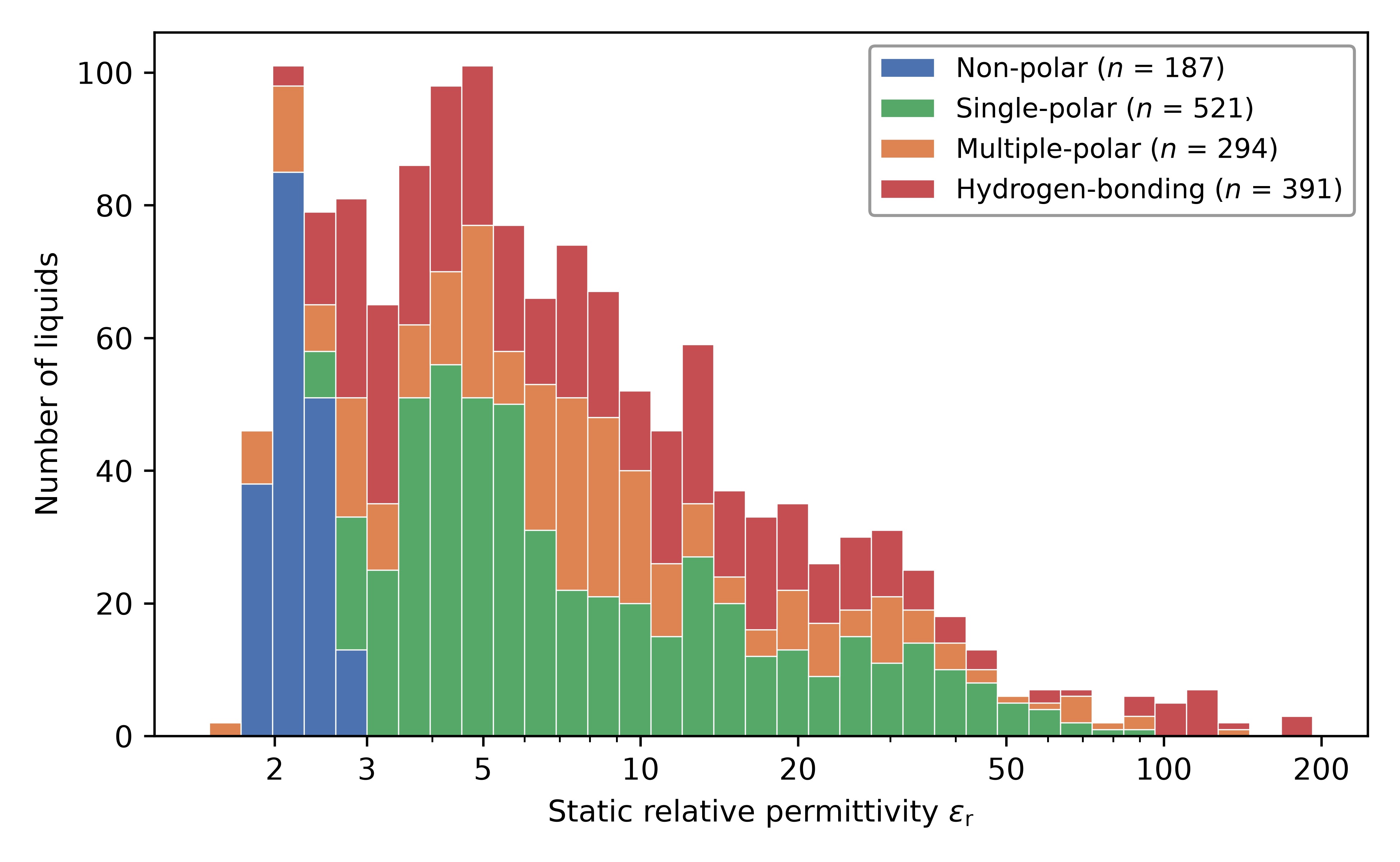}
  \caption{Distribution of the static relative permittivity $\varepsilon_\mathrm{r}$ of the 1393 liquids in
  the dataset, stacked by molecular class. Logarithmic bins.}
  \label{fig:eps_hist}
\end{figure}

According to the classification used by \craft (Section~\ref{sec:parameterization}), the dataset contains 187 non-polar, 521 single-polar, 294 multiple-polar and 391 hydrogen-bonding liquids. Figure~\ref{fig:eps_hist} shows the distribution of the permittivities. They span more than two orders of magnitude, from 1.54 (octafluoropropane) to 183 (\textit{N}-methylformamide), with a median of 5.6. The non-polar liquids form a narrow peak between 1.8 and 3.0. Single-polar, multiple-polar and hydrogen-bonding liquids cover the range from about 2 to beyond 100. Almost all values above 100 belong to hydrogen-bonding \textit{N}-alkylamides, the only exception being the mesoionic 3-methylsydnone ($\varepsilon_\mathrm{r} = 144$).

\subsection{Computational workflow}
\label{sec:workflow}

\begin{figure}[htbp]
  \centering
  \begin{tikzpicture}[
      node distance=6mm and 6mm,
      box/.style={draw, rounded corners=2pt, align=center, font=\small,
                  minimum height=10mm, text width=40mm, inner sep=3pt},
      every edge/.style={draw, -{Stealth[length=2mm]}},
      >={Stealth[length=2mm]}]
    \node[box] (smi) {SMILES};
    \node[box, below=of smi] (conf) {RDKit ETKDGv3 conformers\\diastereomers, polar-H variants};
    \node[box, below=of conf] (xtb) {GFN2-xTB/ALPB\\optimization, single point,\\qRRHO free energy $G_j$};
    \node[box, below=10mm of xtb] (mu) {polar-fragment vectors\\$\vect{\mu}_{i,j}$, $\vect{\mu}_{\mathrm{rest},j}$\\$\Rightarrow\overline{\mu_{\mathrm{eff}}^2}$};
    \node[box, left=of mu] (alpha) {atomic polarizabilities\\$\overline{\alpha}_a$, scales $s_t$\\$\Rightarrow\varepsilon_\infty$ (Clausius--Mossotti)};
    \node[box, right=of mu] (vm) {group-contribution\\molar volume\\$V_m(T)$};
    \node[box, below=10mm of mu] (eps) {Onsager--Kirkwood--Fr\"ohlich\\$\Rightarrow\varepsilon_r$};
    \draw[->] (smi) -- (conf);
    \draw[->] (conf) -- (xtb);
    \draw[->] (xtb) -- (mu);
    \draw[->] (xtb.south) -- ++(0,-3mm) -| (alpha.north);
    \draw[->] (smi.east) -| (vm.north);
    \draw[->] (alpha.south) |- (eps.west);
    \draw[->] (mu) -- (eps);
    \draw[->] (vm.south) |- (eps.east);
  \end{tikzpicture}
  \caption{Workflow of \craft. Conformer ensembles and their electronic properties are computed once per
    molecule; the fitted scale factors enter only in the last two rows.}
  \label{fig:workflow}
\end{figure}
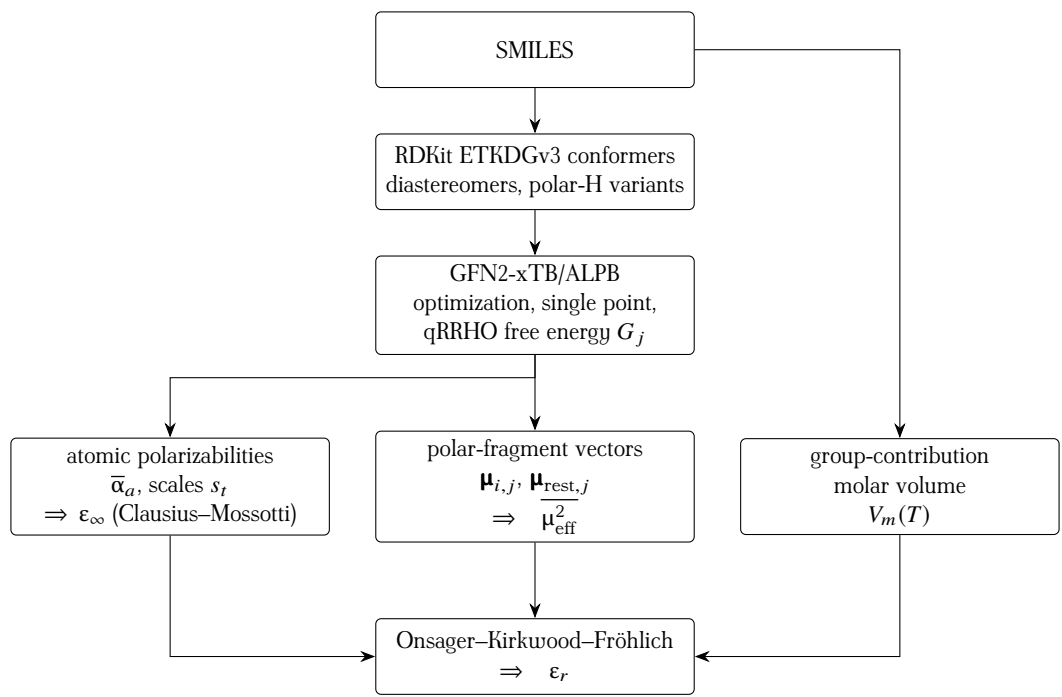

Figure~\ref{fig:workflow} summarizes the workflow from a SMILES string to $\varepsilon_r$. Starting from the canonical SMILES, up to 100 conformers are embedded with the ETKDGv3 distance-geometry method of RDKit \cite{landrum_rdkit,riniker_better_2015,wang_improving_2020} (heavy-atom RMSD pruning threshold 0.125\,\AA). Stereo elements whose configuration is not specified in the SMILES are treated as follows: if they lie within two bonds of a polar atom, so that their relative configuration can change the molecular dipole, every distinct relative configuration is sampled as a separate diastereomer ensemble. The 100 embedding seeds are shared between these ensembles. For molecules with at least two rotatable terminal OH, SH or NH$_2$ groups, every embedded conformer is supplemented with two variants with random polar-hydrogen torsions and one variant in which the polar hydrogens are directed towards feasible intramolecular acceptors. This samples both intramolecularly hydrogen-bonded and open arrangements.

All conformers are optimized with GFN2-xTB \cite{bannwarth_gfn2-xtb_2019,bannwarth_extended_2021} (xtb~6.7.1, tight convergence) in the ALPB implicit-solvent model \cite{ehlert_robust_2021}. A single-point calculation on the optimized geometry provides the molecular dipole, the GFN2 partial charges and atomic dipole moments, and the static atomic polarizabilities $\alpha_a(0)$ of the charge-dependent D4 dispersion model \cite{caldeweyher_generally_2019}. The sum of the point-charge and atomic-dipole contributions is checked against the molecular dipole printed by xtb for every conformer. Duplicates are removed by energy-ordered leader clustering: a conformer is assigned to an existing cluster if its dipole moment differs by less than 0.2\,D and its all-atom symmetry-corrected RMSD (carbon-bound hydrogens removed, reflections allowed) \cite{meli_spyrmsd_2020} is below 0.5\,\AA. For the lowest-energy member of each cluster, a numerical Hessian in the same continuum gives the free energy $G_j$ in the modified rigid-rotor--harmonic-oscillator (qRRHO) approximation at 298.15\,K, with a free-rotor treatment of modes below 50\,cm$^{-1}$ \cite{grimme_supramolecular_2012}. Chiral conformers are stored together with their mirror image, which has the same energy and polarizability and a reflected dipole vector. On average, a molecule is represented by 72 conformers.

Conformer ensembles were generated in 23 ALPB solvents and in the gas phase. \craft uses diethyl ether as continuum, which gave the lowest cross-validated error (Section~\ref{sec:solvent}). The molar volume at 298.15\,K is taken from the atom-type group contributions of Bouteloup and Mathieu \cite{bouteloup_predicting_2019}. The conformer workflow requires only modest computational resources. With up to eight single-threaded xtb
processes running in parallel on a laptop, the median wall time per liquid is 10\,s, and 78\,\% of the 1393 liquids are completed within one minute and 96\,\% within ten minutes. Only large, flexible molecules such as triglycerides require up to about 3\,h, mainly because of the numerical Hessians for the
free energies, so the complete dataset is processed in about 45\,h of wall time (more details in the Supporting Information). \craft is provided as the pip-installable Python package \texttt{craft-epsilon} (source code at \url{https://github.com/simonmb/craft-epsilon}; a copy is included in the Supporting Information), which runs the complete workflow from a SMILES string or a compound name to $\varepsilon_r$ (\texttt{craft-epsilon "CCO"} on the command line, or \texttt{epsilon("CCO")} in Python), provided the xtb program is installed. Names are resolved to cross-checked structures with MoleculeResolver \cite{muller_how_2025}. All fitted parameters are defined in this package.

\subsection{Model formulation}
\label{sec:model}

The static relative permittivity is calculated from the Onsager--Kirkwood--Fr\"ohlich relation \cite{onsager_electric_1936,kirkwood_dielectric_1939,frohlich_theory_1958},

\begin{equation}
\frac{
\left(2\varepsilon_r+\varepsilon_\infty\right)
\left(\varepsilon_r-\varepsilon_\infty\right)
}{
\varepsilon_r
\left(\varepsilon_\infty+2\right)^2
}
=
\frac{
N_A\,\overline{\mu_{\mathrm{eff}}^2}
}{
9\varepsilon_0 k_B T V_m(T)
}
\equiv A,
\label{eq:onsager_pure}
\end{equation}

whose physical root is
\begin{equation}
\varepsilon_r=\frac{\beta+\sqrt{\beta^2+8\varepsilon_\infty^2}}{4},
\qquad
\beta=\varepsilon_\infty+A\left(\varepsilon_\infty+2\right)^2 .
\label{eq:onsager_root}
\end{equation}

Here, $\varepsilon_\infty$ is the high-frequency relative permittivity, $N_A$ is Avogadro's constant, $\varepsilon_0$ is the vacuum permittivity, $k_B$ is Boltzmann's constant, $T$ is the temperature of the experimental data point, and $V_m(T)$ is the molar volume. The Kirkwood correlation factor is not calculated explicitly; orientational correlation is absorbed into the effective dipole $\overline{\mu_{\mathrm{eff}}^2}$ defined below.

The group-contribution molar volume is available at $T_0=298.15$~K only. For data points measured at other temperatures it is corrected to first order with the isobaric thermal expansion coefficient $\alpha_V=\left(\partial \ln V/\partial T\right)_p$,

\begin{equation}
V_m(T)=V_m(T_0)\left[1+\alpha_V\left(T-T_0\right)\right].
\label{eq:molar_volume}
\end{equation}

A single, fixed value $\alpha_V=1.0\times10^{-3}$~K$^{-1}$ is used, representative of organic liquids at ambient conditions, whose expansion coefficients typically lie between about $0.7\times10^{-3}$ and $1.4\times10^{-3}$~K$^{-1}$. $\alpha_V$ is not fitted. Most data points (1177 of 1393) were measured within 5~K of $T_0$, where the correction is negligible. The results are insensitive to the chosen value: varying $\alpha_V$ between $0.5\times10^{-3}$ and $2.0\times10^{-3}$~K$^{-1}$ changes the five-fold cross-validated AAD by less than 0.05 (2.38--2.44; Table~S3 of the Supporting Information). A value fitted to the permittivity data would, in addition, partly compensate for the temperature dependence of hydrogen-bond association that is not represented in the model, rather than describe the thermal expansion of the liquid.

\subsubsection{Calculation of the effective dipole}
The xTB/ALPB molecular dipole of conformer $j$ is decomposed exactly into vectors associated with the identified polar groups $i$ and a residual contribution,

\begin{equation}
\vect{\mu}^{\mathrm{xTB}}_j
=
\vect{\mu}_{\mathrm{rest},j}
+
\sum_i
\vect{\mu}_{i,j}.
\label{eq:xtb_dipole_decomposition}
\end{equation}

The vector of polar group $i$ combines the GFN2 partial charges and atomic dipole moments of its atoms,

\begin{equation}
\vect{\mu}_{i,j}
=
\kappa_D
\sum_{a\in\mathcal{A}_i}
q_{a,j}
\left(
\vect{r}_{a,j}
-
\vect{r}_{i,j}^{\,\mathrm{anchor}}
\right)
+
\sum_{a\in\mathcal{A}_i}
\vect{p}_{a,j},
\label{eq:polar_group_dipole}
\end{equation}

where the set $\mathcal{A}_i$ contains the atoms of polar group $i$, including hydrogens bonded to its heteroatoms; $q_{a,j}$ (in $e$), $\vect{p}_{a,j}$ and $\vect{r}_{a,j}$ (in \AA) are the partial charge, atomic dipole moment and Cartesian position of atom $a$; and $\kappa_D=4.80320471257~\mathrm{D\,(e\,\text{\AA})^{-1}}$. The atomic dipoles, which xtb reports in $e\,a_0$, are converted to debye with $2.541746473~\mathrm{D\,(e\,a_0)^{-1}}$. The anchor $\vect{r}_{i,j}^{\,\mathrm{anchor}}$ is the mean position of the heavy atoms bonded to the group (of all bonded atoms if there is no heavy neighbour, of the group itself if it has no neighbour). Anchoring at the attachment atoms makes the group vector independent of the group's position in the molecule: the group's net charge, placed at the anchor, remains in $\vect{\mu}_{\mathrm{rest},j}$.

The effective liquid-phase dipole of conformer $j$ is

\begin{equation}
\vect{\mu}_{\mathrm{eff},j}
=
b\,\vect{\mu}_{\mathrm{rest},j}
+
\sum_i c_{k(i)}\,\vect{\mu}_{i,j},
\label{eq:effective_dipole}
\end{equation}

where $b$ is a fitted remainder scale and $c_{k(i)}$ is the fitted scale of the parameter class $k(i)$ of polar group $i$ (a chemical family in the polar stage, a fragment type in the hydrogen-bonding stage; see Section~\ref{sec:parameterization}). For nonpolar liquids, which contain no polar groups, $\vect{\mu}_{\mathrm{eff},j}=c_\mathrm{np}\,\vect{\mu}^{\mathrm{xTB}}_j$ with a single fitted scale $c_\mathrm{np}$. Because each group vector is scaled before the vector sum, the model can change both the magnitude of the molecular dipole and the degree of cancellation or reinforcement between polar groups; $c_k^2$ can be read as an effective group correlation factor.

The orientational input is the Boltzmann-averaged \emph{mean-square} effective dipole,

\begin{equation}
\overline{\mu_{\mathrm{eff}}^2}
=
\sum_j w_j
\left\lVert
\vect{\mu}_{\mathrm{eff},j}
\right\rVert^2.
\label{eq:mean_square_effective_dipole}
\end{equation}

The conformer weights use the ALPB qRRHO free energies $G_j$ at $T_0=298.15$~K, the temperature of the conformer ensembles,

\begin{equation}
w_j=
\frac{1}{n_\mathrm{d}}\,
\frac{\exp\left[-\left(G_j-\min_{l\in\mathcal{D}(j)} G_l\right)/RT_0\right]}
{\sum_{l\in\mathcal{D}(j)}\exp\left[-\left(G_l-\min_{l'\in\mathcal{D}(j)} G_{l'}\right)/RT_0\right]},
\qquad
\sum_jw_j=1,
\label{eq:boltzmann_weights}
\end{equation}

where $\mathcal{D}(j)$ is the diastereomer ensemble containing conformer $j$ and $n_\mathrm{d}$ is the number of diastereomer ensembles of the molecule. Diastereomer ensembles are averaged with equal weight (an equimolar mixture); because chiral conformers are stored together with their mirror image, every enantiomeric pair enters twice and every achiral conformer once.

\subsubsection{Calculation of \texorpdfstring{$\varepsilon_\infty$}{the high-frequency permittivity}}
The high-frequency relative permittivity $\varepsilon_\infty$ is calculated with the Clausius--Mossotti relation,

\begin{equation}
\frac{
\varepsilon_\infty-1
}{
\varepsilon_\infty+2
}
=
\frac{
N_A\alpha_{\mathrm{mol}}^{\mathrm{SI}}
}{
3\varepsilon_0V_m(T)
},
\label{eq:clausius_mossotti}
\end{equation}

where $\alpha_{\mathrm{mol}}^{\mathrm{SI}}$ is the fitted molecular polarizability in SI units.
For atom $a$ in conformer $j$, the xTB/ALPB calculation provides an atom-resolved polarizability $\alpha_{a,j}$. The Boltzmann-averaged atomic polarizability, with the weights of Equation~\eqref{eq:boltzmann_weights}, is

\begin{equation}
\overline{\alpha}_a=
\sum_jw_j\alpha_{a,j}.
\label{eq:alpha_average}
\end{equation}

Each atom is assigned an atom type $t(a)$ (element and hybridization, with a ring/acyclic flag for C, N, O, P, S; halogens, H and Si by element). Types differing only in the ring/acyclic flag share one scale, and the fitted molecular polarizability volume is

\begin{equation}
A_\alpha=
\kappa_\alpha\sum_a
s_{t(a)}
\overline{\alpha}_a,
\qquad
\kappa_\alpha=0.14818471~
\text{\AA}^3\,\mathrm{a_0^{-3}},
\label{eq:scaled_alpha}
\end{equation}

where $s_t$ is a fitted atom-group scale factor ($s_{\mathrm{H}}=1$ throughout). The quantity used in Equation~\eqref{eq:clausius_mossotti} is then

\begin{equation}
\alpha_{\mathrm{mol}}^{\mathrm{SI}}
=
4\pi\varepsilon_0
\left(A_\alpha\times10^{-30}\right),
\label{eq:alpha_SI}
\end{equation}

where $A_\alpha$ is in \AA$^3$.
\subsection{Assumptions}
\label{sec:limitations}

The interpretation of \craft is subject to the following assumptions, ordered from the
molecular input to the liquid:
\begin{enumerate}
  \item \textbf{Fixed molecular state.} Each molecular input is treated as a fixed tautomeric and protonation state. Molecules with several relevant tautomers lie outside the domain of \craft unless these states are treated explicitly.
  \item \textbf{Conformer populations.} Conformer populations are calculated from GFN2-xTB qRRHO free energies of isolated molecules in an implicit ALPB continuum at 298.15\,K. Their temperature dependence and any changes of the conformational equilibria caused by the surrounding liquid are neglected.
  \item \textbf{Additive, isotropic polarizability.} The electronic polarizability is assumed to be additive and isotropic at the level of the atom-group representation. Explicit anisotropy and intermolecular polarization if present are assumed to be described by xTB.
  \item \textbf{Effective dipole scaling.} The electronic-structure dipoles are scaled with fitted fragment factors and a factor for the remainder dipole. These corrections are effective liquid-phase parameters. They do not represent a unique physical mapping from the molecular dipole in the continuum to the dipole in the liquid.
  \item \textbf{Mean-field closure.} The Onsager--Kirkwood--Fr\"ohlich relation is used as a mean-field closure with $g_\mathrm{K}=1$. Pairwise dipolar correlations and a molecular Kirkwood factor are not calculated explicitly.
  \item \textbf{No explicit liquid structure.} Coordination numbers, hydrogen-bond networks and long-range orientational correlations are not sampled directly. Their average effect, together with the correlations neglected by the closure, is absorbed by the fitted dipole scales.
  \item \textbf{Molar volume.} The molar volume is obtained from the group-contribution method of Bouteloup and Mathieu \cite{bouteloup_predicting_2019}, which is formally a 298.15\,K model. It is extrapolated to the measurement temperature with a fixed volumetric thermal expansion coefficient, which constitutes an additional approximation.
\end{enumerate}

The dipole moments entering \craft are computed with the semi-empirical GFN2-xTB method, both in the geometry optimization and in the subsequent single-point calculation (Section~\ref{sec:workflow}). Benchmark studies show that accurate molecular dipole moments require considerably higher levels of theory, namely hybrid or double-hybrid density functionals with large, diffuse basis sets \cite{hickey_benchmarking_2014,hait_how_2018}, and the dipoles of GFN2-xTB cannot be expected to reach this accuracy. The consequences are limited by the way the dipoles are used. The fitted scales of the polar fragments and of the remainder (Eq.~\eqref{eq:effective_dipole}) absorb a systematic, group-specific over- or underestimation of the dipoles, so that no accurate gas-phase dipole is required as such. Errors that are not uniform within a chemical family, or that depend on the relative orientation and the distance of neighbouring polar groups, are not removed by the scales. They are the likely origin of part of the remaining errors for multiple-polar liquids with closely spaced groups (Section~\ref{sec:results}). The same applies to the conformer populations, which depend on semi-empirical free energies. A higher-level dipole calculation, for example with hybrid DFT for the lowest-energy conformers of the Boltzmann ensemble, could reduce these errors, but at a computational cost that conflicts with the aim of a computationally inexpensive workflow.

\subsection{Parameterization}
\label{sec:parameterization}
Molecules are assigned a class: molecules with at least one hydrogen-bond donor atom (O--H, N--H or S--H; RDKit \texttt{CalcNumHBD}) form the hydrogen-bonding class. For all other molecules, prioritized SMARTS rules identify exclusive polar fragments, in analogy to the automatic fragmentation of molecules into the groups of group-contribution methods \cite{muller_flexible_2019} (each atom belongs to at most one fragment; heteroatoms not matched by any rule form an individual fragment of type ``other''). No fragment gives nonpolar, one gives single-polar, and two or more give multiple-polar. All parameters are bounded ($0\le s_t,c_k\le10$, $0\le c_\mathrm{np},b\le5$) and fitted by least squares on $\varepsilon_r$, with the regularization term of Equation~\eqref{eq:loss} in the polar stage; \craft has 41 parameters in total ($\alpha_V$ is fixed and not counted).

The model is fitted in three stages:
\begin{enumerate}
\item \textbf{Nonpolar stage (3 parameters).}
Atom-group polarizability scales $s_t$ for groups present in at least five nonpolar liquids (here $\mathrm{C_{sp^2}}$ and $\mathrm{C_{sp^3}}$) and the nonpolar dipole scale $c_\mathrm{np}$ are fitted to the nonpolar liquids. All other atom groups keep $s_t=1$.

\item \textbf{Single-polar and multiple-polar stage (21 parameters).}
The nonpolar-stage scales are retained. Scales of the remaining atom groups present in at least 25 polar molecules (11 groups), nine chemical-family scales $c_f$ (carbonyl, halogen, nitro, nitrogen, oxygen, phosphorus, silicon, sulfur, other) and the remainder scale $b$ are fitted jointly to the single-polar and multiple-polar liquids. Every polar group uses the scale of its family, $c_{k(i)}=c_{f(i)}$. The free atom-group scales are regularized toward unity:
\begin{equation}
\mathcal{L}=
\sum_m
\left(
\varepsilon_{r,m}^{\mathrm{model}}-
\varepsilon_{r,m}^{\mathrm{exp}}
\right)^2+
30
\sum_{t\in\mathcal{T}_{\mathrm{free}}}
(s_t-1)^2.
\label{eq:loss}
\end{equation}
The regularization applies only to the free atom-group scales, which can correlate with the family scales; the family scales and $b$ are not regularized.

\item \textbf{Hydrogen-bonding stage (17 parameters).}
The atom-group scales of the polar stage are held fixed. One scale $c_k$ per fragment type $k$ (aromatic and alicyclic variants of a fragment merged) and a separate remainder scale $b_{\mathrm{HB}}$ are fitted to the hydrogen-bonding liquids. Fragment types present in fewer than two hydrogen-bonding molecules are pooled into a common rare-fragment scale. With the present data this gives 15 fragment scales, one rare scale and $b_{\mathrm{HB}}$.
\end{enumerate}
In the fit to the full dataset, three parameters of the hydrogen-bonding stage converge to their lower bound of zero: the remainder scale $b_{\mathrm{HB}}$ and the scales of the chloro and ether fragments in hydrogen-bonding molecules. The fitted values of all parameters are given in the Python package \texttt{craft-epsilon} in the Supporting Information.
\subsection{Evaluation}
\label{sec:metrics}

For experimental relative permittivities
$\varepsilon_i^{\mathrm{exp}}$ and the corresponding predictions
$\varepsilon_i^{\mathrm{pred}}$, the average absolute deviation is

\begin{equation}
\mathrm{AAD}
=
\frac{1}{N}
\sum_{i=1}^{N}
\left|
\varepsilon_i^{\mathrm{pred}}
-
\varepsilon_i^{\mathrm{exp}}
\right|,
\end{equation}

the root-mean-square error is

\begin{equation}
\mathrm{RMSE}
=
\sqrt{
\frac{1}{N}
\sum_{i=1}^{N}
\left(
\varepsilon_i^{\mathrm{pred}}
-
\varepsilon_i^{\mathrm{exp}}
\right)^2
},
\end{equation}

and the coefficient of determination is

\begin{equation}
R^2
=
1-
\frac{
\sum_i
\left(
\varepsilon_i^{\mathrm{pred}}
-
\varepsilon_i^{\mathrm{exp}}
\right)^2
}{
\sum_i
\left(
\varepsilon_i^{\mathrm{exp}}
-
\overline{\varepsilon^{\mathrm{exp}}}
\right)^2
}.
\end{equation}

The statistics are reported for the whole dataset and separately for the nonpolar, single-polar, multiple-polar and hydrogen-bonding classes, in order to identify the chemical regimes responsible for systematic deviations; $N$ and $\overline{\varepsilon^{\mathrm{exp}}}$ then refer to the respective subset. Besides the in-sample fit, the predictive accuracy is assessed by stratified five-fold cross-validation: the molecules of each class are distributed randomly over five folds, each fold is predicted by a model fitted to the remaining four, and all three stages are refitted for every fold. As the folds are drawn at random within each class, the cross-validation tests the prediction of liquids within the chemical space of the data set and not the extrapolation to entirely new functionalities. The continuum used for the conformer ensembles (Section~\ref{sec:solvent}) was selected on the basis of the cross-validated error, so the reported cross-validated accuracy is slightly optimistic with respect to this choice.

\section{Results and Discussion}
\label{sec:results}

\subsection{Overall results}
\begin{figure}[htbp]
  \centering
  \includegraphics[width=0.6\textwidth]{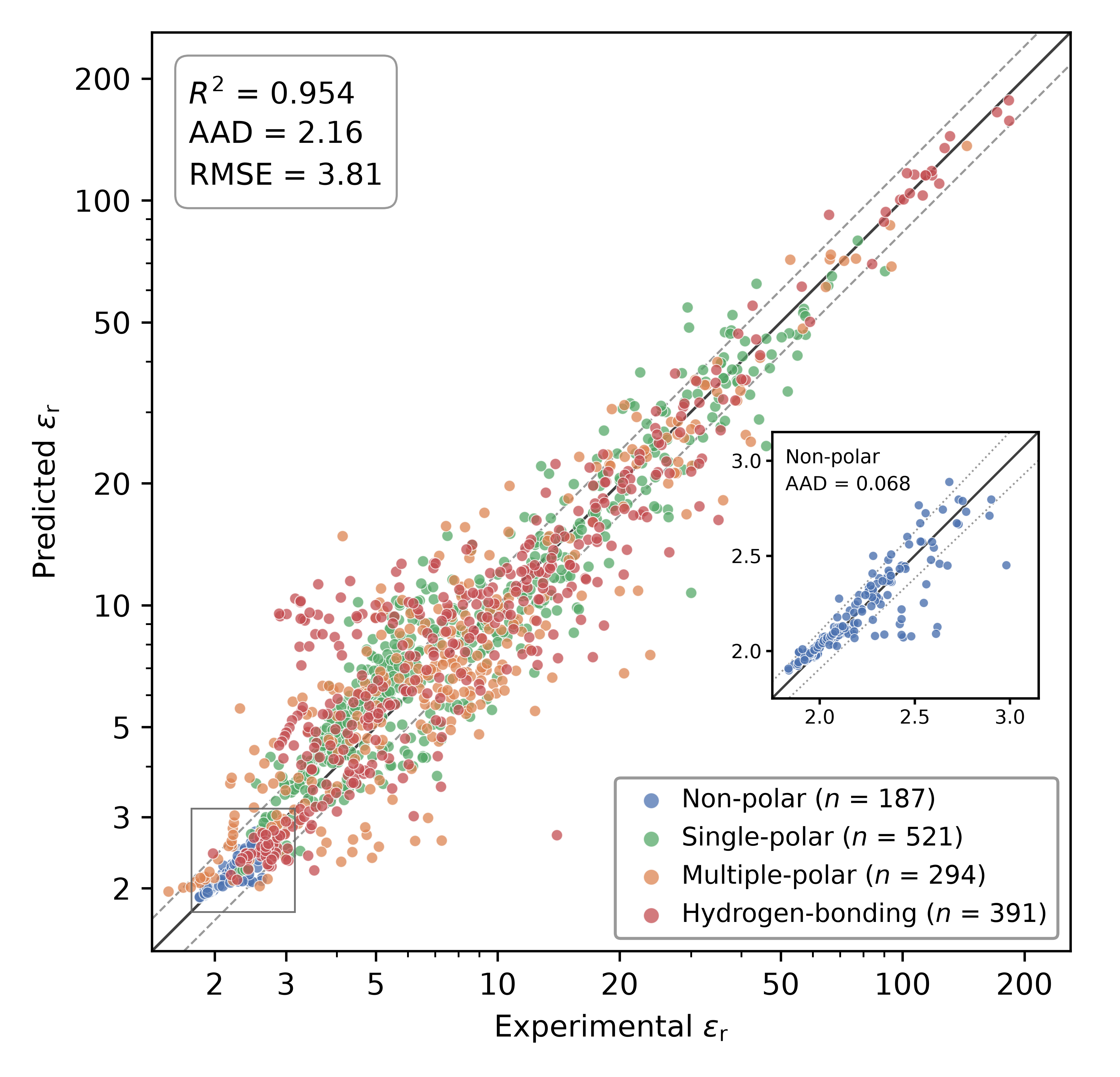}
  \caption{Logarithmic plot of the predicted versus experimental static relative permittivity of the 1393 liquids for \craft with conformer ensembles from diethyl ether (in-sample fit, logarithmic axes). Colours denote the molecular classes. Solid line: parity; dashed lines: $\pm 20\,\%$ deviation. The inset enlarges the non-polar liquids (grey rectangle) on linear axes; dotted lines: $\pm 5\,\%$ deviation.}
  \label{fig:parity_41p}
\end{figure}

With conformer ensembles from diethyl ether, \craft reproduces the relative permittivity of the 1393 liquids with an in-sample AAD of 2.16 (RMSE 3.81, $R^2=0.954$; Figure~\ref{fig:parity_41p}). In five-fold cross-validation the AAD increases to 2.40 (RMSE 4.65, $R^2=0.932$). The moderate gap of about 11\,\% in AAD between in-sample and cross-validated errors indicates that the 41 parameters are relatively well determined by the data and that the model transfers to liquids not included in the fit. Because the error grows with the magnitude of $\varepsilon_r$, the absolute statistics are dominated by the strongly polar and hydrogen-bonding liquids; the classes are therefore discussed separately in Table~\ref{tab:results_41p}.

For non-polar liquids the model is very accurate (cross-validated AAD 0.069, median relative error 1.8\,\%), and 84\,\% of these liquids are reproduced within $\pm 5\,\%$ (inset in Figure~\ref{fig:parity_41p}). The comparatively low coefficient of determination ($R^2=0.77$) does not indicate a poor description but reflects the narrow range of $\varepsilon_r$ in this class (about 1.8 to 3.0), which is dominated by $\varepsilon_\infty$. The inset shows two systematic groups of deviations. Aromatic hydrocarbons without a large permanent dipole, such as benzene, toluene and the xylenes, are overestimated by 5 to 10\,\%. Because their permittivity is determined almost entirely by $\varepsilon_\infty$, this deviation reflects the common polarizability scale of aromatic and olefinic sp$^2$ carbon atoms, whose fitted value is a compromise between both types of hydrocarbons. Conversely, a group of alkenes, alkynes and dienes is underestimated by up to 20\,\%. For terminal alkynes and phenylacetylene, the small but non-negligible dipole moment is not fully captured by the single non-polar dipole scale. For several branched trisubstituted alkenes, for example 2,6-dimethyl-2-heptene and 2,5-dimethyl-2-hexene, the measured permittivities would require dipole moments of
about 0.6 to 1\,D, which are implausible for such hydrocarbons. These values are more likely to reflect experimental uncertainty than a deficiency of the model.

Single-polar and multiple-polar liquids are described with cross-validated AADs of 2.41 and 2.86, respectively. Two systematic effects limit the accuracy in these classes. First, the predictions are slightly compressed: liquids with low permittivity tend to be overestimated and those with high permittivity underestimated. Second, all polar groups of a chemical family share one scale factor, while individual fragment types within a family deviate in opposite directions (for example chloro versus bromo and iodo groups, or esters versus carbonates). Cyclic and acyclic representatives of the same group (e.g.\ cyclic and open-chain sulfones) are also not distinguished. For multiple-polar liquids, molecules whose polar groups are separated by only one or two bonds show the largest errors. Although xTB captures the electronic coupling of neighbouring groups, scaling the partially cancelling group dipoles with different family factors, together with conformer populations computed in a continuum rather than in the liquid itself, amplifies small deviations in the net dipole.

The hydrogen-bonding class is reproduced with $R^2=0.949$ in cross-validation, the highest of all classes, which reflects the wide range of permittivities in this class (up to about 180 for \textit{N}-methylamides). The absolute errors are the largest (AAD 3.15). The difference between in-sample and cross-validated errors is most pronounced, because this stage has the highest ratio of class-specific parameters to molecules (17 parameters for 391 liquids, compared with 21 parameters for the 815 aprotic polar liquids). The fitted fragment scales are consistent with the known association behaviour: scales above unity for amides and alcohols correspond to chain-like hydrogen-bonded structures that align molecular dipoles, whereas carboxylic acids, which form non-polar cyclic dimers, obtain small scales. The fitted remainder scale of this stage lies at its lower bound of zero, i.e.\ the permittivity of hydrogen-bonding liquids is described entirely by their fragment dipoles. Liquids with two donor atoms, such as glycols, oligoethylene glycols, amino alcohols, diamines and dithiols, are systematically underestimated (median relative deviation $-13\,\%$ in cross-validation, and up to about $-50\,\%$ for oligoethylene glycols, resorcinol and 1,2-diamines). A single fragment scale presumably cannot represent the three-dimensional hydrogen-bond networks of these liquids. With ensembles from the hydrogen-bond-accepting continuum diethyl ether, liquids with three or more donor groups (polyols) show no systematic bias (median relative deviation 0\,\%). The hydrogen-bonding class as a whole is described better than with non-accepting continua such as the gas phase, chloroform or benzene (cross-validated AAD 3.15 compared with 3.35--3.38; Section~\ref{sec:solvent}), whereas the underestimation of the two-donor liquids persists.

The high-frequency permittivity $\varepsilon_\infty$ of the model is not fitted to refractive indices, but can be compared with the square of the experimental refractive index $n_D$ for the 792 liquids for which a value is available in the compilation of Bouteloup and Mathieu \cite{bouteloup_improved_2018}. The compilation is based on the catalog of Tokyo Chemical Industry (Table~S5 of the Supporting Information). For non-polar liquids, $\varepsilon_\infty$ agrees with $n_D^2$ within a mean absolute deviation of 0.05 (2.4\,\%). For the polar classes the deviation is larger (mean absolute deviation 0.14--0.45), and $\varepsilon_\infty$ is on average higher than $n_D^2$, presumably because the polarizability scales, which are fitted to the permittivity, also absorb part of the orientational contribution. The largest deviations occur for polyhalogenated compounds. As $\varepsilon_\infty$ is an effective quantity, this comparison is a plausibility check and not a validation of the polarizabilities.

\begin{table}[htbp]
\centering
\caption{Performance of \craft with conformer ensembles from GFN2-xTB/ALPB in diethyl ether. In-sample statistics refer to the fit to all 1393 liquids; cross-validated (CV) statistics to stratified five-fold cross-validation, in which every fold is predicted by a model refitted to the remaining four.}
\label{tab:results_41p}
\begin{tabular}{lrcccccc}
\toprule
 & & \multicolumn{3}{c}{In-sample} & \multicolumn{3}{c}{5-fold CV} \\
\cmidrule(lr){3-5}\cmidrule(lr){6-8}
Class & $N$ & AAD & RMSE & $R^2$ & AAD & RMSE & $R^2$ \\
\midrule
nonpolar          &  187 & 0.068 & 0.113 & 0.782 & 0.069 & 0.115 & 0.773 \\
single-polar      &  521 & 2.296 & 3.908 & 0.903 & 2.411 & 4.229 & 0.886 \\
multiple-polar    &  294 & 2.605 & 4.219 & 0.929 & 2.855 & 4.936 & 0.903 \\
hydrogen-bonding  &  391 & 2.638 & 4.254 & 0.973 & 3.146 & 5.907 & 0.949 \\
\midrule
all liquids       & 1393 & 2.158 & 3.814 & 0.954 & 2.396 & 4.651 & 0.932 \\
\bottomrule
\end{tabular}
\end{table}

\subsection{Influence of the ALPB solvent}
\label{sec:solvent}

The overall accuracy depends only weakly on the implicit solvent used to generate the conformer ensembles (Tables~S1 and S2 of the Supporting Information): with the model refitted for each of the 23 ALPB solvents and the gas phase, the cross-validated AAD spans only 2.40--2.51. The 14 best-ranked environments lie within 0.07 of each other, which is comparable to the variation expected from a single random fold assignment. No systematic trend with solvent polarity is observed, because the fitted scale factors absorb most of the uniform changes of dipole moments and polarizabilities induced by the continuum; the solvent acts mainly through the relative conformer populations. Resolved by class, the sensitivity increases with the polarity and hydrogen-bonding ability of the liquids, from a spread of 0.0014 in AAD for nonpolar to 0.27 for hydrogen-bonding liquids. For the latter, the five best continua are all hydrogen-bond acceptors (diethyl ether, DMF, DMSO, THF, 1,4-dioxane) and improve the AAD by about 0.2 relative to the gas phase and to non-accepting continua such as chloroform, benzene or dichloromethane (3.15--3.19 versus 3.35--3.42). The reason is that the solvent-specific hydrogen-bond correction of ALPB stabilizes conformers whose OH and NH groups point towards the continuum rather than forming intramolecular hydrogen bonds. This mimics a liquid in which intramolecular hydrogen bonds are largely replaced by intermolecular ones. This preference is shared by a whole group of chemically related solvents and is therefore robust, whereas the lead of an individual solvent over its closest competitors (0.01--0.05 in AAD) should not be overinterpreted. Combining the best solvent of each class does not improve the model (cross-validated AAD 2.56), since the classes share atom-group, family and remainder scales that cannot reconcile ensembles with systematically different dipole moments. All conformer ensembles are therefore generated in one common continuum, diethyl ether.

\subsection{Effective dipole moments}
\label{sec:dipoles}

\begin{figure}[htbp]
  \centering
  \includegraphics[width=\textwidth]{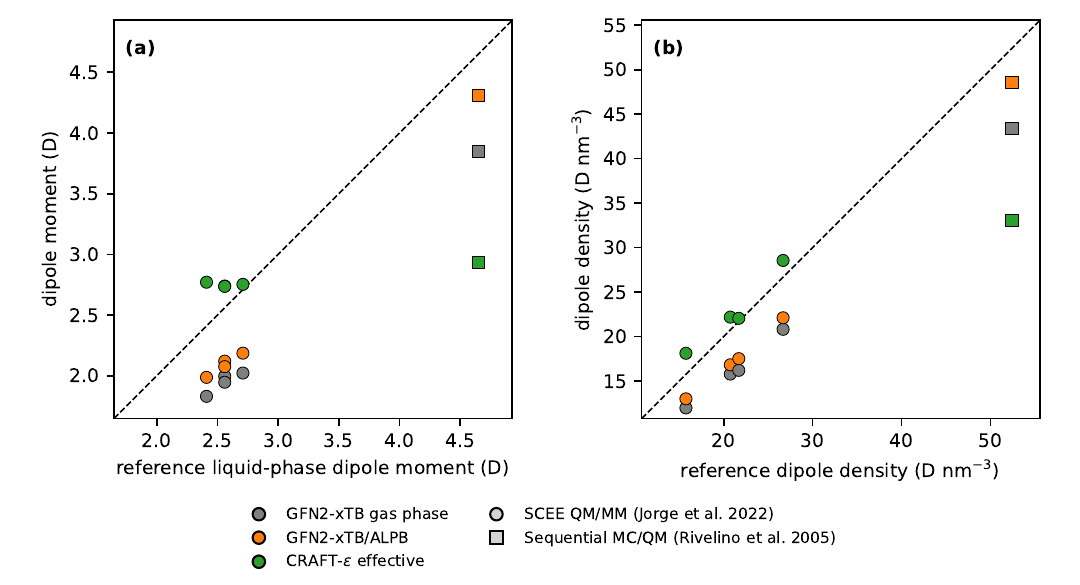}
  \caption{Parity plots of (a) the molecular dipole moment and (b) the dipole density $\mu N_\mathrm{A}/V_\mathrm{m}$ of four alcohols (circles) and acetonitrile (squares) against liquid-phase reference values from QM/MM simulations (alcohols \cite{jorge_dipole_2022}; acetonitrile \cite{rivelino_electronic_2005}). Grey: GFN2-xTB in the gas phase; orange: GFN2-xTB/ALPB (diethyl ether); green: effective dipole of \craft after scaling. All values are Boltzmann-averaged over the conformer ensemble; the dipole density uses the model molar volume at the measurement temperature. Dashed line: parity.}
  \label{fig:dipole_parity}
\end{figure}

Because the dipoles enter \craft only in scaled form, it is instructive to compare the resulting effective dipole moments $\sqrt{\overline{\mu_\mathrm{eff}^2}}$ (Eq.~\eqref{eq:effective_dipole}) with condensed-phase dipole moments from higher-level simulations. Such values are available for few liquids; here, four alcohols (ethanol, 1-propanol, 2-propanol, \emph{tert}-butanol) from self-consistent electrostatic-embedding QM/MM simulations at the B3LYP level \cite{jorge_dipole_2022} and acetonitrile from sequential Monte Carlo/quantum mechanics simulations at the MP2 level \cite{rivelino_electronic_2005} are considered. Figure~\ref{fig:dipole_parity} compares these reference values, as dipole moments (a) and as dipole densities $\mu N_\mathrm{A}/V_\mathrm{m}$ (b), with the Boltzmann-averaged root-mean-square dipole of the GFN2-xTB gas-phase ensemble, of the GFN2-xTB/ALPB ensemble used here, and with the effective dipole of the final model (Table~S6 of the Supporting Information). The dipole density is the quantity that governs the orientational contribution to the permittivity (Eq.~\eqref{eq:onsager_pure}), where the dipole enters through $\overline{\mu_\mathrm{eff}^2}/V_\mathrm{m}$.

The unscaled xTB dipoles underestimate the liquid-phase references. In the gas phase the deficit is 17\,\% for acetonitrile and 22--25\,\% for the alcohols; the ALPB continuum raises the dipoles and reduces the deficit to 7\,\% and 17--19\,\%, respectively. Using ALPB-based ensembles brings the dipoles closer to the liquid-phase reference than gas-phase calculations, as expected for a continuum that mimics the polarization of the molecule by its environment, although it recovers only part of the enhancement found in the simulations. The effective dipoles of \craft deviate from the references in a different manner for the two groups of liquids. For the hydrogen-bonding alcohols the scaled dipoles (2.74--2.77\,D) are close to the references (2.41--2.71\,D), with deviations of +2 to +15\,\%, and are thus closer to the reference than the unscaled ones. For acetonitrile, a non-hydrogen-bonding liquid with a strongly polar group, the scaled dipole (2.93\,D) lies 37\,\% below the reference (4.65\,D) and is clearly farther from it than the ALPB value (4.31\,D). The dipole densities show the same ordering.

This behaviour is a consequence of what the scale factors represent. They are fitted to the permittivity, not to the molecular dipole. They absorb every effect that is not contained in the dipole of an isolated molecule in a continuum: the packing of the molecules, the orientational correlation between neighbouring dipoles (the Kirkwood factor, which is not calculated explicitly but absorbed into the effective dipole, Section~\ref{sec:model}), and the specific interactions of hydrogen bonds, in addition to the polarization of the molecule by its surroundings. In hydrogen-bonding liquids, chain-like association tends to align neighbouring dipoles, and the polarization of the hydroxyl group by its hydrogen-bonded neighbours enhances the dipole at the same time. The fitted scales are accordingly larger than one (Section~\ref{sec:results}), and the effective dipoles of the alcohols approach the simulated condensed-phase values. In non-hydrogen-bonding polar liquids, in contrast, the scaling damps the dipole below the molecular value. A plausible reading is that antiparallel dipole correlations and the mean-field character of the Onsager reaction field, which the scales also have to absorb, outweigh the polarization enhancement in these liquids. The agreement for the alcohols should therefore not be read as a validation of the scaled dipoles as molecular properties. The reduced dipole of acetonitrile is not a statement that its liquid-phase dipole is small: both are effective parameters of a permittivity model. The reference values themselves also depend on the partitioning of the electron density between molecules and on the level of theory, so that agreement within about 10\,\% cannot be resolved. The comparison nevertheless shows that the unscaled dipoles are systematically too small relative to condensed-phase values, and that the fitted scales correct them in the hydrogen-bonding liquids, where the effect is largest, but overcompensate for polar aprotic liquids with large dipole moments.

\subsection{Influence of conformer ensemble and energy}
\begin{table}[htbp]
\centering
\caption{Effect of the conformer description on \craft (in-sample fit, 1393 liquids). The model is refitted for each variant with the same 41 parameters. \emph{Ensemble}: Boltzmann-weighted ensemble of GFN2-xTB/ALPB(diethyl ether) conformers with qRRHO free-energy weights at 298.15\,K (on average 72 conformers per molecule). \emph{Lowest $G$}: only the conformer with the lowest qRRHO free energy. \emph{Lowest $E$}: only the conformer with the lowest electronic energy. AAD and RMSE are absolute deviations in $\varepsilon_\mathrm{r}$. The best value in each row is shown in bold.}
\label{tab:lowest_conformer}
\begin{tabular}{l ccc ccc ccc}
\toprule
 & \multicolumn{3}{c}{Ensemble} & \multicolumn{3}{c}{Lowest $G$} & \multicolumn{3}{c}{Lowest $E$} \\
\cmidrule(lr){2-4}\cmidrule(lr){5-7}\cmidrule(lr){8-10}
Class & AAD & RMSE & $R^2$ & AAD & RMSE & $R^2$ & AAD & RMSE & $R^2$ \\
\midrule
Overall        & \textbf{2.158} & \textbf{3.814} & \textbf{0.9545} & 2.574 & 4.861 & 0.9261 & 2.576 & 4.893 & 0.9251 \\
Non-polar      & \textbf{0.068} & \textbf{0.113} & 0.7816 & \textbf{0.068} & \textbf{0.113} & 0.7815 & \textbf{0.068} & \textbf{0.113} & \textbf{0.7829} \\
Single-polar   & 2.296 & 3.908 & 0.9029 & 2.275 & 3.909 & 0.9028 & \textbf{2.224} & \textbf{3.868} & \textbf{0.9048} \\
Multiple-polar & \textbf{2.605} & \textbf{4.219} & \textbf{0.9291} & 3.590 & 6.266 & 0.8435 & 3.609 & 6.266 & 0.8435 \\
Hydrogen-bonding & \textbf{2.638} & \textbf{4.254} & \textbf{0.9734} & 3.405 & 5.856 & 0.9496 & 3.468 & 5.987 & 0.9474 \\
\bottomrule
\end{tabular}
\end{table}
Table~\ref{tab:lowest_conformer} shows the importance of conformational averaging. If the Boltzmann-weighted ensemble is replaced by a single lowest-energy conformer, the overall AAD increases from 2.16 to 2.57 and $R^2$ decreases from 0.955 to 0.926, although the number of parameters stays the same. Whether the conformer is selected by free energy or by electronic energy makes almost no difference.

The deterioration is confined to the classes in which the molecular dipole moment depends on the conformation. For multiple-polar liquids, the AAD increases by almost 40\,\% and $R^2$ decreases from 0.93 to 0.84. In these molecules the resulting dipole is set by the relative orientation of the polar groups, for example gauche and anti arrangements in glycol ethers, diesters or dinitriles. A single conformer often represents an arrangement in which the group dipoles largely cancel, so it cannot reproduce the ensemble-averaged $\langle\mu^2\rangle$. Figure~S1 of the Supporting Information shows this for 1,2-dibromoethane, hexanedinitrile and 2,3-butanediol: in each of them the most populated conformer has an effective dipole moment of at most 0.06\,D, whereas the Boltzmann-averaged value is 0.82, 3.41 and 3.19\,D, respectively. Hydrogen-bonding liquids with several donor groups behave similarly. In 2,3-butanediol, the most populated conformer is an extended arrangement without intramolecular hydrogen bond and with a vanishing dipole, whereas the conformers with an intramolecular O--H$\cdots$O hydrogen bond have effective dipoles of about 4\,D. A single conformer is therefore not representative of the ensemble of such liquids.

For non-polar and single-polar liquids, in contrast, a single conformer is as good as the ensemble or even marginally better. Here the dipole moment is dominated by one rigid group, or it is negligible, and is therefore nearly independent of the conformation. The same trends are obtained in five-fold cross-validation, where the overall AAD increases from 2.40 for the ensemble to 2.81 and 2.86 for the lowest-$G$ and lowest-$E$ conformers. A conformational ensemble is therefore essential for liquids with several polar or hydrogen-bonding groups, whereas a single conformer is sufficient for non-polar and single-polar liquids.

\subsection{Comparison with the model of Bouteloup and Mathieu}
\label{sec:comparison_bouteloup}

Bouteloup and Mathieu \cite{bouteloup_predicting_2019} predict $\varepsilon_r$ from the SMILES string alone. The molar volume
$V_m$ and the refractive index $n_D$ are obtained from atom-type group contributions, and the orientational term
$g\mu^2$ is a sum of transferable polar-fragment contributions. These are corrected for the presence of
hydrogen atoms adjacent to some fragments and for the interaction between neighbouring polar fragments, which decays
with the topological distance between them. All three quantities enter the Onsager--Kirkwood--Fr\"ohlich equation. The model
uses 74 fragment contributions, of which six depend on the number of adjacent hydrogen atoms and thus carry two
parameters each, together with 11 interaction parameters and one decay length $\tau$, for a total of 92 parameters
fitted to permittivity data. The molar refraction of each of 47 atom types adds a further 47 parameters for $n_D$,
giving 139 parameters in all. Both models obtain $V_m$ from the same group contributions of Bouteloup and Mathieu, so these are not counted for either model.

\begin{figure}[htbp]
  \centering
  \begin{minipage}[c]{0.55\linewidth}
    \centering
    \includegraphics[width=\linewidth]{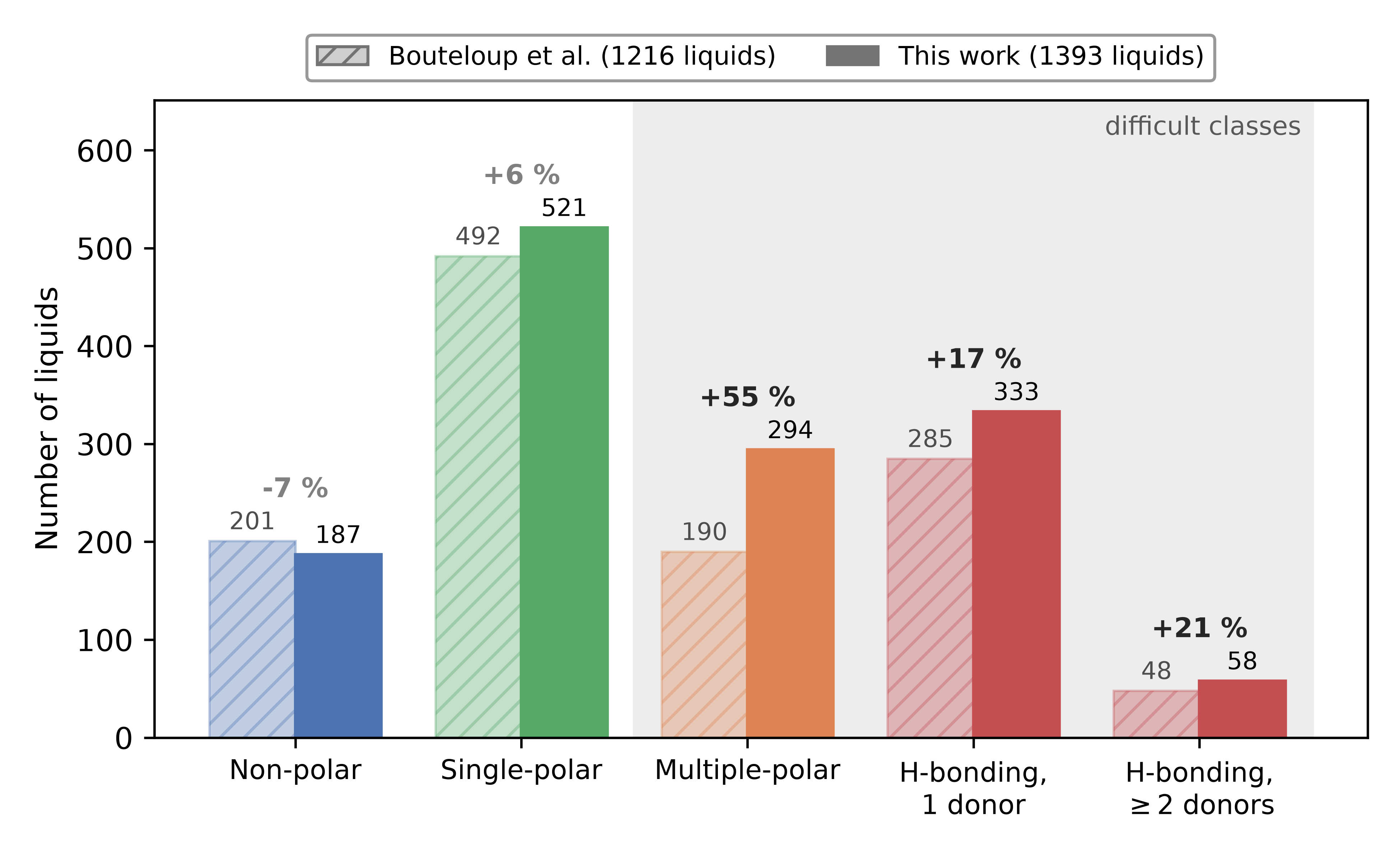}\\[0.2em]
    {\small (a)}
  \end{minipage}\hfill
  \begin{minipage}[c]{0.43\linewidth}
    \centering
    \includegraphics[width=\linewidth]{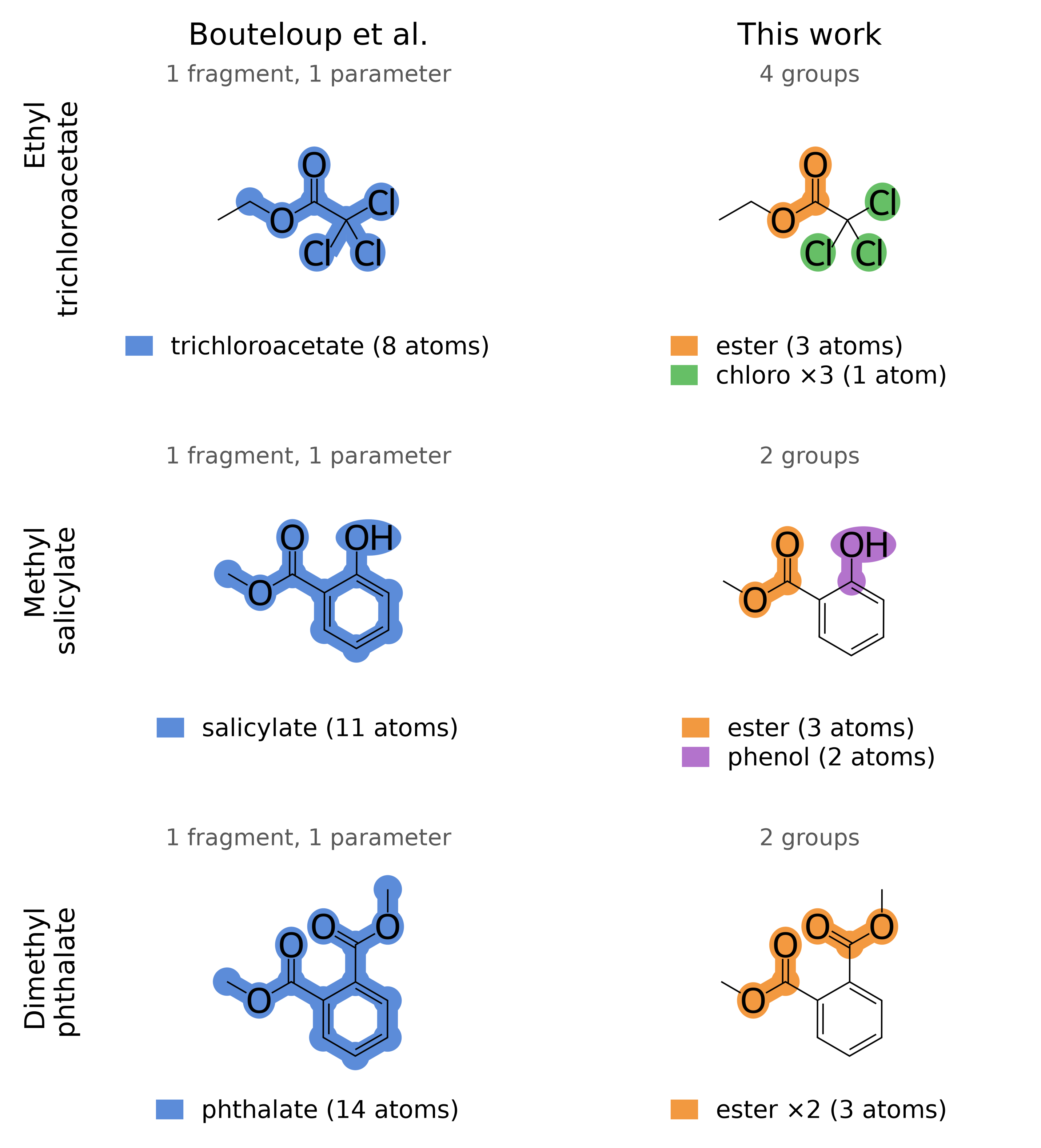}\\[0.2em]
    {\small (b)}
  \end{minipage}
  \caption{Comparison with the model of Bouteloup and Mathieu \cite{bouteloup_predicting_2019}.
    (a) Number of liquids per molecular class in the data collection of Bouteloup and Mathieu (hatched) and in
    the dataset of this work (\craft dataset, solid), both classified with the rule of \craft;
    hydrogen-bonding liquids are split by the number of donor groups. Numbers above the bars give the relative
    difference of the \craft dataset; the grey band marks the classes that are most difficult to describe.
    (b) Polar groups of three example molecules in the model of Bouteloup and Mathieu (left) and in \craft
    (right). Highlighted atoms belong to a polar group; colours distinguish the groups. Each fragment of
    Bouteloup and Mathieu carries one fitted $g\mu^2$ contribution. Atom counts are heavy atoms and, for
    Bouteloup and Mathieu, include the carbon atoms at which the molecule is cut.}
  \label{fig:fragment_comparison}
\end{figure}

The two models differ most in the size of their polar groups (Fig.~\ref{fig:fragment_comparison}b). Bouteloup
and Mathieu cut a molecule only at single bonds between an sp$^3$ carbon and another sp$^3$ or aromatic carbon, so a
polar group stays connected to every conjugated or aromatic unit and to every other polar group in direct contact
with it. In ethyl trichloroacetate, the ester and the three C--Cl groups form a single eight-atom fragment. In methyl
salicylate, the ester, the phenol and the benzene ring form one fragment of eleven atoms, and in dimethyl phthalate
the two esters and the ring form one of fourteen. \craft describes the same molecules with small,
reusable groups of one to three heavy atoms: an ester and three chloro groups, an ester and a phenol, and two esters.
Across the dataset, a fragment of Bouteloup and Mathieu contains 4.3 heavy atoms on average, against 1.8 for a group of
\craft.

Large fragments have an advantage: a single empirical constant absorbs intramolecular effects such as conjugation
with the ring or the intramolecular hydrogen bond of methyl salicylate, which explains the high accuracy of Bouteloup
and Mathieu for the aromatic esters and anilines in their library. The price is that every combination of polar groups,
and every substitution pattern on a ring, becomes a new fragment type. The three isomeric dichlorobenzenes, for
instance, give three distinct fragments, and 2- and 4-nitrophenol give two. Covering the present dataset would
require 182 distinct fragment types, of which only 74 are parametrized; 73 of the 185 molecules the model cannot
evaluate fail on a merged fragment of eight or more atoms. In \craft, the same ester, phenol or chloro
group recurs in all of these molecules, its scale is shared by every molecule containing it (through nine
fragment-family scales for aprotic polar liquids and per-group scales for hydrogen-bonding liquids), and the
differences between isomers enter through the computed conformer dipoles.

The two datasets also differ in composition (Fig.~\ref{fig:fragment_comparison}a). Both contain a similar number of
non-polar and single-polar liquids, but the present dataset contains 55\,\% more multiple-polar liquids (294 compared
with 190) and about 20\,\% more hydrogen-bonding liquids, including those with two or more donor groups. These are
the classes in which neighbouring polar groups interact and in which the molecular dipole depends most strongly on
the conformation, and therefore the classes that are most difficult to describe. They are also the classes in which
the fragment library of Bouteloup and Mathieu is least complete: their model cannot evaluate 34\,\% of the
multiple-polar and 13 to 16\,\% of the hydrogen-bonding liquids of the present dataset, compared with 7\,\% of the
single-polar and none of the non-polar liquids. The comparison on shared molecules below therefore excludes mainly
the difficult part of the present dataset.

The model of Bouteloup and Mathieu can only be evaluated for a molecule if every polar fragment it contains has a parameter. This is the case for 1208 of the 1393 molecules in the present dataset. The remaining 185 molecules contain a fragment with missing parameters. All comparisons below use these 1208 shared molecules. The model of Bouteloup and Mathieu is reported in two forms: with the published parameters, and with the 92 permittivity parameters refitted to the same data following their parametrization procedure, with the 47 refraction parameters kept at their published values. Both the refitted model of Bouteloup and Mathieu and \craft are cross-validated with the same stratified five-fold split, so each held-out molecule is predicted by both models after training on identical data. In cross-validation, 62 held-out molecules contain a fragment that occurs in fewer than three training molecules of the other four folds, the minimum required by Bouteloup and Mathieu to assign a parameter. They therefore cannot be predicted by the refitted model, and the cross-validated statistics refer to the 1146 molecules both models predict.

\begin{table}[tp]
  \centering
  \caption{Comparison with the model of Bouteloup and Mathieu \cite{bouteloup_predicting_2019} on the 1208 molecules
    both models can evaluate. $N_\mathrm{par}$ is the number of fitted parameters: for Bouteloup and Mathieu, the 92
    parameters of the dipolar term plus the 47 atom-type parameters of the refractive index; the group-contribution
    molar volume used by both models is not counted. The cross-validated (CV) statistics are from stratified
    five-fold cross-validation on the 1146 molecules both models predict.}
  \label{tab:comparison_bouteloup}
  \begin{tabular}{lrrrrrrr}
    \toprule
     & & \multicolumn{3}{c}{Fit ($n = 1208$)} & \multicolumn{3}{c}{5-fold CV ($n = 1146$)} \\
    \cmidrule(lr){3-5}\cmidrule(lr){6-8}
    Model & $N_\mathrm{par}$ & AAD & RMSE & $R^2$ & AAD & RMSE & $R^2$ \\
    \midrule
    Bouteloup and Mathieu (paper)     & 139         & 1.49 & 3.72 & 0.959 & 1.42\textsuperscript{a} & 3.48\textsuperscript{a} & 0.956\textsuperscript{a} \\
    Bouteloup and Mathieu (refitted)  & 139         & 1.42 & 3.06 & 0.972 & 1.54 & 3.41 & 0.958 \\
    \craft (this work)            & 37\textsuperscript{b} & 2.03 & 3.57 & 0.962 & 2.05 & 3.65 & 0.951 \\
    \bottomrule
  \end{tabular}

  \smallskip
  \begin{minipage}{0.95\linewidth}\footnotesize
    \textsuperscript{a}~Published parameters evaluated on the 1146 cross-validation molecules without refitting;
    these parameters were fitted to literature data that largely overlap with the present dataset, so the values are
    not out-of-sample.
    \textsuperscript{b}~Atom-group polarizability scales are only fitted for groups present in enough molecules; on the
    shared subset 37 of the 41 parameters of the full model are fitted (34--35 per cross-validation fold).
  \end{minipage}
\end{table}

\begin{table}[bp]
  \centering
  \caption{Cross-validated errors by molecular class (stratified five-fold CV, identical folds for both models).}
  \label{tab:comparison_bouteloup_classes}
  \begin{tabular}{lrrrrrrr}
    \toprule
     & & \multicolumn{3}{c}{Bouteloup and Mathieu (refitted)} & \multicolumn{3}{c}{\craft (this work)} \\
    \cmidrule(lr){3-5}\cmidrule(lr){6-8}
    Class & $n$ & AAD & RMSE & $R^2$ & AAD & RMSE & $R^2$ \\
    \midrule
    Non-polar       & 184 & 0.100 & 0.163 & 0.546 & \textbf{0.069} & \textbf{0.116} & \textbf{0.772} \\
    Single polar    & 456 & \textbf{1.22} & \textbf{2.77} & \textbf{0.945} & 2.21 & 3.93 & 0.890 \\
    Multiple polar  & 180 & \textbf{1.57} & \textbf{3.55} & \textbf{0.856} & 2.46 & 4.11 & 0.807 \\
    H-bonding       & 326 & 2.78 & 4.81 & 0.965 & \textbf{2.71} & \textbf{4.01} & \textbf{0.976} \\
    \midrule
    All             & 1146 & \textbf{1.54} & \textbf{3.41} & \textbf{0.958} & 2.05 & 3.65 & 0.951 \\
    \bottomrule
  \end{tabular}
\end{table}

Table~\ref{tab:comparison_bouteloup} shows that the model of Bouteloup and Mathieu reaches a lower average error than \craft, with a cross-validated AAD of 1.54 against 2.05. The gap in RMSE is much smaller (3.41 against 3.65), and the coefficients of determination are nearly equal (0.958 against 0.951). This advantage requires 139 parameters, almost four times the 37 of \craft. With about 970 training molecules per fold, the model of Bouteloup and Mathieu has one parameter per seven molecules, against one per 26 for \craft. The in-sample statistics in Table~\ref{tab:comparison_bouteloup} refer to 1208 molecules and the cross-validated ones to 1146. Evaluated on the same 1146 molecules, the RMSE increases from fit to cross-validation by about 0.5 for both models (2.90 to 3.41 for the refitted model of Bouteloup and Mathieu, 3.14 to 3.65 for \craft), and the AAD by 13\,\% and 9\,\%, respectively. Owing to the size of the dataset, the larger number of parameters of Bouteloup and Mathieu therefore does not lead to markedly stronger overfitting; the advantage of \craft lies in its parameter economy and coverage rather than in a smaller fit-to-prediction gap.

The class-resolved errors in Table~\ref{tab:comparison_bouteloup_classes} show where each approach is strongest. The model of Bouteloup and Mathieu is clearly more accurate for aprotic polar liquids, where its AAD is lower by 0.99 (single polar) and 0.89 (multiple polar) than that of \craft. These are the compounds for which the model was designed: every polar fragment carries its own empirical $g\mu^2$ contribution, which absorbs both the fragment dipole and its typical orientational correlation. \craft uses nine fragment-family scales instead, which cannot follow such fragment-specific differences.

\craft is more accurate for non-polar liquids (AAD 0.069 against 0.100, $R^2$ 0.77 against 0.55), where
the small orientational contribution follows from the computed molecular dipole and the polarizability rather than
from a near-zero fragment constant. It is also more accurate for hydrogen-bonding liquids (RMSE 4.01 against 4.81,
$R^2$ 0.976 against 0.965). These have the largest permittivities and the strongest association, and the present
model resolves their dipole over conformers instead of treating it as a sum of fixed fragment constants.

Finally, \craft applies to every molecule for which conformers can be generated. The model of Bouteloup
and Mathieu cannot evaluate 185 of the 1393 molecules (13\,\%) with its published parameters, and a further 62 held-out
molecules in cross-validation, because their polar fragments lack a parameter. \craft thus achieves
comparable $R^2$ and RMSE with about a quarter of the parameters and complete coverage of the dataset, whereas the
model of Bouteloup and Mathieu remains the more accurate choice for the aprotic polar compounds its fragment library
covers.

\section{Conclusions}\label{Conclusions}

\craft, a conformer-resolved molecular model for the static relative permittivity of
neutral pure liquids, has been developed and evaluated on 1393 liquids spanning
$\varepsilon_r = 1.5$--$183$. Conformer ensembles, molecular dipoles and
atom-resolved polarizabilities are obtained from GFN2-xTB/ALPB calculations.
Electronic polarization enters through scaled atomic polarizabilities and the
Clausius--Mossotti relation. Orientational polarization enters through the
Boltzmann-averaged mean-square dipole, assembled from scaled polar-fragment
vectors, and the Onsager--Kirkwood--Fr\"ohlich relation. \craft has
41 parameters, fitted in three stages: 3 for nonpolar, 21 for
aprotic polar and 17 for hydrogen-bonding liquids.

With conformer ensembles from ALPB(diethyl ether), the model reproduces the
dataset with an in-sample AAD of 2.16 (RMSE 3.81, $R^2 = 0.954$). In stratified
five-fold cross-validation, the AAD increases only to 2.40 (RMSE 4.65,
$R^2 = 0.932$), so the parameters are well determined and transfer to liquids
not used in the fit. Nonpolar liquids are described within a cross-validated
AAD of 0.069. Hydrogen-bonding liquids, which include the highest
permittivities, reach $R^2 = 0.949$.

Conformational averaging is essential for molecules whose dipole depends on
the relative orientation of several polar or hydrogen-bonding groups.
Replacing the ensemble by a single lowest-energy conformer increases the
cross-validated AAD from 2.40 to 2.81, and the in-sample AAD of multiple-polar
liquids by almost 40\,\%. The choice of implicit solvent affects the overall
accuracy only weakly (cross-validated AAD 2.40--2.51 across 24 environments).
For hydrogen-bonding liquids, however, hydrogen-bond-accepting continua
consistently perform best, as they suppress intramolecular hydrogen bonds that
are not representative of the liquid.

Compared with the fragment-based model of Bouteloup and Mathieu, the present
model reaches a similar cross-validated RMSE (3.65 versus 3.41) and $R^2$
(0.951 versus 0.958) with about a quarter of the parameters (37 versus 139 on
the shared subset). It also covers every liquid in the dataset, whereas 13\,\%
remain outside the fragment library of the reference model. The reference
model remains more accurate for aprotic polar liquids. \craft is
more accurate for nonpolar and hydrogen-bonding liquids.

The remaining errors are systematic and attributable to identifiable
approximations. Family-wide scales cannot separate fragment types within a
family, nor cyclic from acyclic groups. Closely spaced polar groups amplify
small errors in partially cancelling dipoles. Liquids with two donor atoms are
underestimated by a median of 13\,\%, since a single fragment scale cannot
represent their three-dimensional hydrogen-bond networks. Resolving these
effects, and the temperature dependence of association, are the most direct
routes to further improvement. Because the model requires only the molecular
structure, it provides a basis for predicting the composition- and
temperature-dependent permittivity of solvent mixtures and electrolyte
solutions, where dielectric data are often unavailable. With a permittivity
now predicted at ambient conditions, the approaches of Rueben \textit{et al.}
\cite{rueben_predicting_2023} or Maribo-Mogensen \textit{et al.}
\cite{maribo-mogensen_modeling_2013} can be used to predict its temperature
and pressure dependence.


\Contributions{
S.M.: conceptualization, methodology, software, investigation, formal analysis, visualization, writing -- original draft, writing -- review \& editing.
}

\Acknowledgments{
Funded by the Deutsche Forschungsgemeinschaft (DFG) - Project number 503327656.
}

\DataAvailability{
The Python package \texttt{craft-epsilon}, which predicts $\varepsilon_r$ from a SMILES string or a compound name and
includes all fitted parameters, is available at \url{https://github.com/simonmb/craft-epsilon} and can be installed
with \texttt{pip install craft-epsilon}; a copy is included in the Supporting Information. The Supporting Information
also includes the solvent study (Tables~S1 and S2) and the sensitivity to the thermal expansion coefficient
(Table~S3). The GFN2-xTB/ALPB conformer ensembles of all liquids in the 23 implicit solvents and the gas phase,
including geometries, energies, qRRHO free energies, partial charges, atomic dipole moments and atomic
polarizabilities, are available on Zenodo at \url{https://doi.org/10.5281/zenodo.23485587}.
The code used to create the dielectric data set from the original data sets, including all filtering and exclusion
steps, is given in the Supporting Information.
}

\section*{Declaration of generative AI use}
During the preparation of this work the author used Claude (Anthropic) and ChatGPT (OpenAI) to obtain feedback on the structure of the manuscript, to restructure and copy-edit sections of the text, to draft the conclusions and the description of the computational workflow from the author's source code, to format the Supporting Information and the workflow figure, and to re-run the author's existing analysis scripts to check reported statistics. All scientific content, including the model, the code, the data, the analysis and its interpretation, was developed by the author. After using this tool, the author reviewed and edited the content as needed, verified all numerical results and references, and takes full responsibility for the content of the published article.

\EndMatter

\end{document}